\documentclass[12pt]{article}

\usepackage{graphicx}

\def\beq{\begin{eqnarray}}
\def\eeq{\end{eqnarray}}
\def\bea{\begin{eqnarray}}
\def\eea{\end{eqnarray}}
\def\be{\begin{equation}}
\def\ee{\end{equation}}

\def\tev{\, {\rm TeV}}
\def\gev{\, {\rm GeV}}
\def\mev{\, {\rm MeV}}

\def\xfb{\, {\rm fb}}
\newcommand{\gsim}{\lower.7ex\hbox{$\;\stackrel{\textstyle>}{\sim}\;$}}
\newcommand{\lsim}{\lower.7ex\hbox{$\;\stackrel{\textstyle<}{\sim}\;$}}

\begin{document}

\setlength{\baselineskip}{0.25in}

\begin{titlepage}
\noindent
\begin{flushright}
MIFP-07-17 \\
MCTP-07-20 \\
UCI-TR-2007-30\\
\end{flushright}
\vspace{1cm}

\begin{center}
  \begin{Large}
    \begin{bf}
Probing the Green-Schwarz Mechanism \\
at the Large Hadron Collider \\

    \end{bf}
  \end{Large}
\end{center}
\vspace{0.2cm}
\begin{center}
\begin{large}
Jason Kumar$^{1}$, Arvind Rajaraman$^{2}$, James D. Wells$^{3}$ \\
\end{large}
  \vspace{0.3cm}
  \begin{it}
$^{(1)}$Department of Physics, Texas A\&M University, College Station, TX  77843-4242, USA \\
\vspace{0.1cm}
$^{(2)}$Department of Physics, University of California, Irvine, CA  92697, USA \\
\vspace{0.1cm}
$^{(3)}$Michigan Center for Theoretical Physics (MCTP) \\
 Department of Physics, University of Michigan, Ann Arbor, MI  48109, USA \\
\vspace{0.1cm}
\end{it}

\end{center}

\begin{abstract}

 We investigate the
phenomenology of new abelian  gauge bosons, which we denote as $X$
bosons, that suffer a mixed anomaly with the Standard Model, but
are made self-consistent by the Green-Schwarz mechanism.  A
distinguishing aspect of the resulting effective theory is the decay of
$X$ bosons into Standard Model gauge bosons, $X\to ZZ, WW, \gamma
Z$. We compute the production cross-section of the $X$ boson from
vector boson fusion at the Large Hadron Collider.
We study the $pp\to X\to ZZ\to 4l$ signal, and analyze the
prospects of discovery.
We argue that such a discovery could indirectly probe
high energies, even up to the string scale.

\end{abstract}

\vspace{1cm}

\begin{flushleft}
July 2007
\end{flushleft}

\end{titlepage}

\setcounter{footnote}{0} \setcounter{page}{2}
\setcounter{figure}{0} \setcounter{table}{0}


\section*{Introduction}

Many of the most well-motivated ideas for physics beyond the
Standard Model (SM) suggest the existence of new $U(1)$ gauge
bosons. Such gauge bosons occur naturally in $SO(10)$ grand
unified models, extra dimensional models with a hidden sector
brane, and string theoretic models with intersecting branes. If
these gauge bosons are coupled to leptons, they can provide
spectacular and clean signals at colliders.

However, in many models, there need not be a tree-level coupling
between the SM particles and the new $U(1)$ gauge bosons. In
particular, SM-like intersecting brane model (IBM)
constructions~\cite{IBM,IBM2} typically contain relatively large gauge groups of
which the SM is only one sector. In these models, the
SM fermions typically have no couplings to the other
gauge groups.

 In these models, couplings to the hidden sector can be generated at the loop level.
Kinetic mixing can give rise to couplings of the hidden-sector
gauge bosons  to SM states. The quantum corrections that mix the
kinetic terms of the extra gauge boson and hypercharge are
unsuppressed by any mass scales, yielding renormalizable terms
that probe both the ultraviolet (UV) scales and infrared (IR)
scales~\cite{kinmix}.

In this letter we will study a different limit, where the couplings
are dominated by terms from a mixed anomaly (the consistency of
which will be made clear below) between the SM gauge sector and an
exotic $U(1)_X$. In this circumstance, we will have a massive
hidden sector gauge boson $X$ whose main couplings  to the SM
sector are through the gauge fields.   Such a gauge boson can have
distinctive signatures at colliders.

These anomaly terms are of great interest for another reason: they
could indirectly probe  high energy physics, and possibly
even stringy effects. This is because anomalies can be thought of
as both UV and IR effects.  They are clearly visible in the  limit
of low-energy effective field theory which we expect to match to
data, yet their UV character implies that the existence (and
resolution) of these anomalies can often be tied to a stringy
origin.  This stringy origin also implies that in a wide variety
of models, the same types of anomaly effects can be observed, thus
making the anomaly an interesting candidate for a stringy
signature to be observed at colliders.

Gauge anomalies must be cancelled for consistency.  In string
theory this is achieved via the Green-Schwarz mechanism, wherein
closed string couplings yield classical gauge-variant terms whose
variation cancels the anomalous diagrams.  This mechanism can have
phenomenological
consequences~\cite{IBMmixanom,Anastasopoulos:2006cz,anom,Berenstein:2006pk},
and observable
effects at the Large Hadron Collider\footnote{We define the Large
Hadron Collider here to be a $pp$ collider running at $14\tev$
center of mass energy.} (LHC). In particular, we will find that
the measurement of these anomalous couplings can probe the
non-vectorlike couplings of the entire Hagedorn tower of states
that arise in IBM's.

Below, we first review the intersecting brane model setup that
gives motivation to the mixed anomaly couplings of SM gauge bosons
with an exotic $X$ gauge boson\footnote{Our $X$ bosons are not
to be confused with grand unified
theory $X$ bosons.}. We then calculate the effective
vertex that couples the hidden sector $X$ gauge boson to the
visible sector. We will consider the phenomenology of this generic
IBM setup by identifying a simple effective theory description
that obviates the need to further consider the string theory
origin. After having the effective theory description in place, we
compute the decay widths for the hidden sector gauge boson, and
use the results to compute the cross-section for production of $X$
via vector boson fusion. We outline the parameter space that can
be probed by the LHC in the clean four lepton final state mode. We
conclude with a short discussion of our results.

\section*{Intersecting Brane Models and the Effective Theory}

Consider IBM's arising from Type IIA string theory compactified on
an orientifolded CY 3-fold.  The branes in question are
spacetime-filling D6-branes wrapping 3-cycles of the Calabi-Yau.
The SM arises from strings that begin and end on a
certain set of D-branes (the so-called ``visible" sector branes).
Additional D-branes are generally required in order to cancel
RR-tadpoles, or, equivalently, to ensure that all space-filling
charges cancel. These additional D-branes generate gauge groups
beyond the SM (the  ``hidden" sector).

Chiral matter arises from strings stretching between two different
branes or their orientifold images.  For example, if $a$ and $b$
are two stacks of branes with gauge groups $G_a$ and $G_b$ living
on them, the net number of chiral multiplets charged under the
bifundamental of these two groups is counted by the topological
intersection number of these branes, $I_{ab}$. In particular, if
hidden sector branes have non-trivial topological intersection
with visible sector branes, there will be a net number of chiral
multiplets transforming in the bifundamental of SM and hidden
sector gauge groups. Note that these topological intersection
numbers are generically nonzero, as any two 3-cycles will
generally intersect on a 6-manifold.

  Consider a
hidden sector with a diagonal subgroup $U(1)_X$.  Generically,
there will be a non-zero number of chiral multiplets in the
bifundamental of this group and the SM group $SU(2)_L$. These
chiral multiplets will lead to a $[U(1)_X SU(2)_L ^2]$ mixed
anomaly through triangle diagrams with fermions running in the
loop.
This anomaly will be cancelled by the Green-Schwarz
mechanism, which arises from adding to the Lagrangian terms that
are classically gauge-variant, and whose gauge variation cancels
the anomalous triangle diagrams.  There will be two types of such
terms: a generalized Chern-Simons term which couples the gauge
boson of $U(1)_X$ to the $SU(2)_L$ gauge bosons, and a
Peccei-Quinn term which couples the $SU(2)_L$ field strength to
an axion. We may write the Lagrangian as the sum of classically
gauge-invariant and gauge-variant
pieces~\cite{Anastasopoulos:2006cz} as
\bea L_{inv.} &=& -{1\over
2} Tr [F_W ^2] -{1\over 2}(\partial_{\mu} a +M_X X_{\mu})^2 - \bar
\psi \gamma^{\mu} D_{\mu} \psi
\nonumber\\
L_{var.} &=& {D'\over 24 \pi^2} a\, Tr\,F_W \wedge F_W +{Z\over 48
\pi^2} \epsilon^{\mu \nu \rho \sigma} X_{\mu} \Omega_{\nu \rho
\sigma} \eea where ${1\over 2} \partial_{\mu} \Omega_{\nu \rho
\sigma} = Tr[F_W \wedge F_W]$   and the axion has a shift
transformation under $U(1)_X$ of $a \rightarrow a + M_X \lambda$.
In a string theory setting, the Green-Schwarz mechanism arises
automatically.  In the UV limit of the one-loop open string diagram,
the annulus stretches into a closed string which couples at
tree-level to the open strings.
One can also consider the example of a $U(1)_X$ arising in 
the SM sector \cite{Anastasopoulos:2006da}.

These new gauge-variant terms will also contribute
to an effective vertex for a
coupling between the $U(1)_X$ gauge boson and two $SU(2)_L$ gauge
bosons\cite{Anastasopoulos:2006cz,anom}. For a
suitable choice of coefficients $D'$ and $Z$, the
anomalous terms can be cancelled, and the
resulting theory  satisfies the
Ward identity.

Although the divergences and anomaly are cancelled, the triangle
diagram nevertheless contributes an unambiguous finite piece to
the effective vertex operator for an interaction between a
$U(1)_X$ gauge boson and two $SU(2)_L$ vector bosons.
Following \cite{Anastasopoulos:2006cz}, we can write this
effective vertex as \bea \Gamma_{\mu \nu \rho} ^{\alpha} (k_3 ^X,
k_1 ^Z, k_2 ^Z) &=&  t^\alpha [A_1 \epsilon_{\mu \nu \rho \sigma}
k_2 ^{\sigma} + A_2 \epsilon_{\mu \nu \rho \sigma} k_1 ^{\sigma}
+B_1 k_{2\nu }\epsilon_{\mu \rho \sigma \tau} k_2 ^{\sigma} k_1
^{\tau} +B_2 k_{1\nu }\epsilon_{\mu \rho \sigma \tau} k_2
^{\sigma}
k_1 ^{\tau}  \nonumber\\
&\,&+ B_3 k_{2\rho }\epsilon_{\mu \nu \sigma \tau} k_2 ^{\sigma} k_1
^{\tau} +B_4 k_{1\rho }\epsilon_{\mu \nu \sigma \tau} k_2
^{\sigma}
k_1 ^{\tau} \nonumber\\
&\,&+ D{k_{3\mu} \over k_3 ^2}\epsilon_{\nu \rho \sigma \tau} k_2
^{\sigma} k_1 ^{\tau} +Z \epsilon_{\mu \nu \rho \sigma} (k_2
^{\sigma} -k_1 ^{\sigma})] \eea
In this expression, $Z$ arises from the generalized Chern-Simons
terms and $D$ arises from the  Peccei-Quinn terms and from the
one-loop coupling of the fermions to the Goldstone boson (computed
in Lorentz gauge). $A_{1,2}$
and $B_{1,2,3,4}$ are the coefficients of
tensor structures which arise from computation of the triangle
diagrams with three gauge bosons as external legs.  $B_{1,2,3,4}$ are finite and
unambiguously determined by the triangle
diagrams~\cite{Anastasopoulos:2006cz},
\bea B_1(k_1 ,k_2) =-B_4
(k_2 ,k_1) &=& -{\imath g_hg_W^2\over 8\pi^2} \sum_{i}
t_{i}\int_0 ^1
d\alpha d\beta {2\alpha \beta  \over \alpha k_2 ^2 +\beta k_1 ^2
-(\alpha k_2 -\beta k_1)^2 -m_{i} ^2}
\nonumber\\
B_2(k_1 ,k_2) =-B_3 (k_2 ,k_1) &=& -{\imath g_hg_W^2\over 8\pi^2} \sum_{i}t_{i}
\int_0 ^1 d\alpha d\beta {2 \beta (1-\beta) \over \alpha
k_2 ^2 +\beta k_1 ^2 -(\alpha k_2 -\beta k_1)^2-m_{i} ^2}
\eea
where $g_h$ is the gauge coupling of the hidden sector $U(1)_X$,
$g_W$ is the gauge coupling of $SU(2)_L$ and $t_{i}$ encode the charges of the fields
in the loop.
$A_{1,2}$ are  UV cutoff dependent coefficients which
depend on how the diagram is regulated.   Since we have
used Lorentz gauge, the Ward identity takes a simple form:
\bea
\epsilon_3 ^{\mu}  k_1 ^{\nu} \epsilon_2 ^{\rho} \Gamma_{\mu \nu \rho}
= \epsilon_3 ^{\mu} \epsilon_1 ^{\nu} k_2 ^{\rho} \Gamma_{\mu \nu \rho}
=k_3 ^{\mu} \epsilon_1 ^{\nu}  \epsilon_2 ^{\rho} \Gamma_{\mu \nu \rho}=0
\eea
where
$k_1+k_2+k_3=0$. We can remove the divergences in the $A_{1,2}$
coefficients and fix the ambiguity by the redefinition $\tilde A_1
= A_1 +Z$, $\tilde A_2 =A_2 -Z$, yielding \bea \label{vertex}
\Gamma_{\mu \nu \rho} ^{\alpha} (k_3 ^X, k_1 ^W, k_2 ^W) &=&
t^{\alpha} [\tilde A_1 \epsilon_{\mu \nu \rho \sigma} k_2
^{\sigma} + \tilde A_2 \epsilon_{\mu \nu \rho \sigma} k_1
^{\sigma} +B_1 k_{2\nu }\epsilon_{\mu \rho \sigma \tau} k_2
^{\sigma} k_1 ^{\tau} +B_2 k_{1\nu }\epsilon_{\mu \rho \sigma
\tau} k_2 ^{\sigma}
k_1 ^{\tau}  \nonumber\\
&\,&+ B_3 k_{2\rho }\epsilon_{\mu \nu \sigma \tau} k_2 ^{\sigma}
k_1 ^{\tau} +B_4 k_{1\rho }\epsilon_{\mu \nu \sigma \tau} k_2
^{\sigma} k_1 ^{\tau}+ D{k_{3\mu} \over k_3 ^2}\epsilon_{\nu \rho
\sigma \tau} k_2 ^{\sigma} k_1 ^{\tau} ]
\eea
where the Ward identities require
\bea
\tilde A_1 &=& -k_1 \cdot k_2 B_1 - k_1 ^2 B_2 \nonumber\\
\tilde A_2 &=& -k_2 ^2 B_3 -k_1 \cdot k_2 B_4 \nonumber\\
D &=& \tilde A_2 -\tilde A_1 \eea
Note that our effective vertex can now be written
entirely in terms of the computable finite coefficients
$B_{1,2,3,4}$.

The piece of the effective vertex involving the
$\tilde A_{1,2}$ and $B_{1,2,3,4}$ coefficients can
be derived from an effective operator involving generalized
Chern-Simons terms and dimension 6 operators with three
derivatives and 3 gauge fields.  But this operator will not
be gauge invariant, as gauge invariance will only arise once
we include the axionic couplings (in Lorentz gauge the
axions are massless and cannot be integrated out).  One should
be able to write an effective operator which generates the full
vertex in unitary gauge, but gauge invariance will not be manifest
due to the fixed choice of gauge.

\section*{Phenomenology of Mixed Anomalies}

The vertex of eq.~(\ref{vertex}) is very general; it applies to
any scenario where the couplings to a hidden sector are generated
through anomaly diagrams. We will now turn to the phenomenology of
such a coupling.
We will study the particular scenario where $M_{W,Z} \ll M_X$.  In
4 dimensions, one would expect $M_X ^2  \propto
{g_{YM} ^2 \over l_s ^2}\propto
{g_s l_s  \over V_3}$, where $l_s$ is the string length and $V_3$ is
the volume of the 3-cycle wrapped by the hidden
sector brane~\cite{Berenstein:2006pk}.  For a
resonant signal at LHC, we would require $M_X \sim {\cal
O}(\tev)$, which can be accommodated by an appropriate geometry,
string coupling and string scale.

The effective vertex we have found couples the $X$ gauge boson to
two SM gauge bosons. The couplings to different gauge
bosons are model dependent, and are strongly dependent on the
spectrum of the hidden sector. In particular, the relative
couplings of the $X$ boson to gluons and electroweak bosons depend
on the hidden sector spectrum. The phenomenology of $X$ bosons
depends on this relative coupling.

Here, we will  assume that $X$ has couplings only to the electroweak
sector. This is a conservative estimate; if the X boson couples to
gluons, the cross section will be larger than the one we find. The
vertex of eq.~(\ref{vertex}) thus mediates the decays $X$ to $W^+
W^-$, $ZZ$, or $\gamma Z$.  One might expect a $\gamma \gamma$
decay channel as well, but  that this decay is
forbidden when both outgoing particles
are massless. Similarly, $X$ bosons can be produced at the LHC
through the process of vector boson fusion.

We will also assume that the $X$ gauge boson only decays to SM
electroweak gauge bosons.  This assumption would be realized if,
for example, $X$ was lighter than the possible decay products of
the hidden sector.  That scenario is not unreasonable; the chiral
matter of the hidden sector arises from strings stretching between
hidden sector branes, and lives in the bifundamental of the gauge
groups living on both branes.  The mass of the fermions is set by
the symmetry breaking scales of both gauge groups.  So if the
$U(1)_X$ we consider is the hidden sector with the lowest symmetry
breaking scale (likely, if $X$ is in fact the lightest new gauge
boson we see), then the masses of most hidden sector matter will
be dominated by higher symmetry breaking scales, and will thus be
heavier than the $X$.  If this assumption fails, however, then our
signal will be appropriately attenuated.

The rate of these processes is controlled by the magnitude of the
coefficients $B_i$, and in particular is
controlled by $|B_1 - B_2|$. The largest contributions to $|B_1 -
B_2|$ will come from chiral matter running in the loop of the
triangle diagram.  If there are $n$ such multiplets, then one
expects $|B_1 - B_2|$ to scale approximately with $n$. In the
context of an IBM, these $n$ multiplets would arise at $I_{ab}=n$
topological intersections between the $SU(2)_L$ brane stack and
the $U(1)_X$ brane stack.

However, vectorlike matter can
contribute to $|B_1 - B_2|$ as well.  Left-handed and right-handed
multiplets contribute with opposite sign to the integrals which
define $B_{1,2}$.  As such, vectorlike matter pairs of chiral
multiplets give a contribution to $|B_1 - B_2|$ which is
proportional to their mass splitting, and suppressed by two powers
of their mass, which is presumably heavier than the non-vectorlike
matter.  However, in a string scenario there will be an exponentially increasing
Hagedorn spectrum of this matter, which arises from excited string
modes and whose mass spacing will scale at
tree-level like $M_s$.  As a result of the large number of massive
states, the contributions from vectorlike matter can be much
larger than naive mass-scale suppression would suggest.

A large number of exotic states charged under $SU(2)$, as may be
needed to generate a sizeable $|B_1-B_2|$ coefficient, could add
loop contributions to precision electroweak observables that are
incompatible with experimental measurements. However, unlike some
other constraints and the signals we are studying here, a reliable
computation of precision electroweak effects is not possible
without a complete theory, including identification of all
particles, and knowledge of their precise masses and mixings. We
remark that some theories with just one extra multiplet may be
incompatible with the data, and yet some theories with hundreds of
vectorlike multiplets in the TeV range can be compatible without
fine-tuned cancellations. Furthermore, it is possible to have
large couplings to $XVV$ by virtue of very large $U(1)_X$ charges,
but small contributions to precision electroweak observables. We
can only remark here that if $X$ boson signatures become visible
at the LHC, one would need to take into account the precision
electroweak constraints as a necessary guide to build a fully
consistent theory for what has been seen.

The finite contribution of vectorlike matter to the effective
vertex will be highly model specific, depending on the details of
dynamical symmetry breaking as well as supersymmetry breaking.
There is no unique ``string theoretic prediction" for $|B_1 -
B_2|$.  Instead, we will be able to parameterize the
model-dependence of the decay width and cross-section in terms of
a single dimensional parameter $\Lambda_X$ by the definition \be
{
1\over 16}|B_1 - B_2|^2 = {1\over \Lambda_X ^4}, \ee
 From an effective field theory point of view, we can think of
$\Lambda_X$ as the effective scale of a higher-dimensional operator in the
effective Lagrangian which is gauge invariant and couples to $X$
and two $SU(2)_L$ gauge bosons via three-derivative couplings. Our
goal will be to bound the scale $\Lambda_X$ for which detection
will be possible at LHC.


\section*{The  Effective Vector
Boson Approximation}

We will use the narrow width approximation to find the
cross-section for $VV \rightarrow X $ in terms of the decay width
for $X_{\mu}$. We will compute this  using the effective vector
boson approximation (EVBA)~\cite{EVBA,Han:1992hr}, convolving vector boson luminosity
functions  against the hard cross-section:
\begin{eqnarray}
\sigma(pp\to X)&  = &\frac{48\pi^2}{M_X^3}\left\{ 2\Gamma(X\to
W_L^+W_T^-) [\tau_X {\cal L}_{W^+_LW^-_T}(\tau_X)] \right. \nonumber\\
& & ~~~ \left. +\Gamma(X\to Z_LZ_T) [\tau_X{\cal
L}_{Z_LZ_T}(\tau_X)]
  +\Gamma(X\to Z_L\gamma)
[\tau_X{\cal L}_{Z_L\gamma}(\tau_X)] \right\}
\end{eqnarray}
where $\tau_X=M_X^2/s$ with $s$ being the square of the center of
mass energy ($\sqrt{s}=14\,{\rm TeV}$ at the LHC). Here
$\Gamma(X\to W_L^+W_T^-)$,  $\Gamma(X\to Z_LZ_T)$ and $\Gamma(X\to
Z_L\gamma)$ are the partial widths for $X$ to decay to $W$ bosons, $Z$
bosons and $Z\gamma$ respectively. Note that the $X$ decay is
possible only if one of the two outgoing vector bosons is
longitudinally polarized. ${\cal L}_{V_iV_j}(\tau_X)$ are the
effective luminosities for collisions of $V_iV_j$ vector bosons.

The decay widths can be determined from the effective vertex,
and depend only on $M_X$ and $\Lambda_X$.
They can be written as
\bea
\Gamma(X\rightarrow W^+
W^-) & = & (42\mev) \left({1\tev \over \Lambda_X}\right)^4
\left({M_X\over 1\tev}\right)^3 \left(1-{4M_W ^2 \over M_X ^2}\right)^{5/2}  \nonumber\\
\Gamma(X\rightarrow ZZ) & = & (16\mev) \left({1\tev \over
\Lambda_X}\right)^4
\left({M_X\over 1\tev}\right)^3 \left(1-{4M_Z ^2 \over M_X ^2}\right)^{5/2}  \\
\Gamma(X\rightarrow \gamma Z) & = & (4.9\mev) \left({1\tev \over
\Lambda_X}\right)^4 \left({M_X\over 1\tev}\right)^3 \left(1-{M_Z
^2 \over M_X ^2}\right)^{3} \left(1+{M_Z ^2 \over M_X ^2} \right)
\nonumber
\eea

The luminosity of $VV'$ vector boson collisions is determined by
\begin{eqnarray}
{\cal L}_{VV'}(\tau)& = & \int_\tau^1\frac{dy}{y}f_V(y)f_{V'}(\tau/y)
\end{eqnarray}
where  $f_{W_L^+}(y)$, $f_{W_T^-}(y)$, $f_{Z_T}(y)$, $f_{Z_L}(y)$
and $f_\gamma(y)$ are the applicable structure functions for the $W$, $Z$ and
$\gamma$ gauge bosons, in analogy to the quark and gluon structure
functions.  After applying the leading order Callan-Gross relation $F_2(x,Q^2)=2xF_1(x,Q^2)$, a
convenient formulation of the $W$ and $Z$ structure
functions   can be extracted from~\cite{Han:1992hr}:
\begin{equation}
f_{V_T}(x)=\frac{g_V^2}{32\pi^2}\frac{1}{x}\int_x^1 \frac{dy}{y}
F_1^V(y,M_V^2) \left[ 2(y-x)+\frac{x^2}{y}\right] \ln \left( 1+
\frac{sy(y-x)}{M_V^2}\right)
\end{equation}
\begin{equation}
f_{V_L}(x)=\frac{g_V^2}{32\pi^2}\frac{1}{x}\int_x^1 \frac{dy}{y}
F_1^V(y,M_V^2) \left[ 2(y-x)\right]
 \end{equation}
where
\begin{eqnarray}
F_1^{W^+}(x,Q^2) & = & u(x,Q^2)+\bar d(x,Q^2)+\bar s(x,Q^2) \nonumber\\
F_1^{W^-}(x,Q^2) & = & \bar u(x,Q^2)+d(x,Q^2)+s(x,Q^2) \\
F_1^Z(x,Q^2) & = & c_u [u(x,Q^2)+\bar u(x,Q^2)] + c_d
[d(x,Q^2)+\bar d(x,Q^2)+s(x,Q^2)+\bar s(x,Q^2)] \nonumber
\end{eqnarray}
and
\begin{eqnarray}
c_u & = & (T^3_L-2e \sin^2\theta_W)^2+(T^3_L)^2 = \left(
\frac{1}{2}-2\cdot\frac{2}{3}\sin^2\theta_W
\right)^2 + \left( \frac{1}{2}\right)^2 \simeq 0.29 \\
c_d & = & (T^3_L-2e \sin^2\theta_W)^2+(T^3_L)^2 = \left(
-\frac{1}{2}+2\cdot\frac{1}{3}\sin^2\theta_W \right)^2 + \left(
-\frac{1}{2}\right)^2 \simeq 0.37
\end{eqnarray}
For the quark
distribution functions
we use the CTEQ6 set~\cite{cteq}.
Furthermore, in the formulae for $f_V(x)$,  $V_T$ is a transversely polarized vector boson $V$
with mass $M_V$, and $f_{V_T}(x)$ is averaged over the two
transverse polarizations.  $V_L$ is a longitudinally polarized
vector boson $V$. The coefficients $g_V$ are given by $g_W=g\simeq
0.65$ and $g_Z=g/\cos\theta_W\simeq 0.74$. The variable $x$ is the
usual Bjorken variable $x=Q^2/2P\cdot q$, where $P$ is the proton
four-momentum, $q$ is the vector boson four momentum and
$Q^2=-q^2$.  Thus, $x$ is the usual variable of parton
distribution functions.

The photon structure function is obtained
in the standard way of integrating the photon splitting function
over the quarks.  The leading log result is
\begin{equation}
f_\gamma(x,Q^2)= \frac{\alpha}{2\pi}\sum_q \int_x^1 \frac{dy}{y}
P_{\gamma q}\left(\frac{x}{y}\right)[q(y,Q^2)+\bar q(y,Q^2)] \ln
\frac{Q^2}{Q_0^2},
\end{equation}
 where $Q_0=0.25\gev$ is the factorization scale and
\begin{equation}
P_{\gamma q}(z)=Q_q^2\left\{ \frac{1+(1-z)^2}{z}\right\}
\end{equation}
is the $q\to \gamma q$ splitting function.

\section*{Collider Phenomenology}

Although there are several different decay channels for the $X$,
we will primarily
consider the decay $X\rightarrow ZZ \rightarrow 4l$. A heavy $X$ boson
decays to a $ZZ$ pair approximately $25\%$ of the time, and this
channel suffers from the additional small branching fraction for
$B(Z \rightarrow ll)=6.7\%$ (we only consider $e$ and $\mu$ decays). But
this suppression is compensated for by the fact that the resulting
signal is one of the cleanest to measure.

Four-lepton backgrounds from vector boson fusion
have been well studied in the context of LHC Higgs
searches~\cite{Bagger:1995mk,VBFHiggs}.
This is a
particularly interesting production channel, because the two
vector bosons are accompanied by spectator jets from the process
$q \rightarrow qV$. One can remove background events very
efficiently by cutting on these spectator jets~\cite{tagged
jets,Bagger:1995mk}. One demands that events have exactly two
outgoing high $p_T$, high pseudo-rapidity jets obeying the cuts:
\begin{itemize}
\item{$E(j_{tag}>0.8\tev)$} \item{$3.0 < |y(j_{tag})| <5.0$}
\item{$p_T (j_{tag}) >40\gev$}
\end{itemize}
These spectator jets are emitted when the production vector bosons
are generated, and not at the hard scattering process.  As a
result, for a resonance production signal, the amount of signal
which is removed by the jet cuts is largely independent of the
physics of the hard process, including even the energy
scale.
As such, the fraction of signal lost
to the jet cuts is approximately the same as in a Higgs signal,
which
has a $\sim 40 \%$ efficiency~\cite{Bagger:1995mk}.

\begin{figure}[tb]
\centering
\includegraphics[width=15cm]{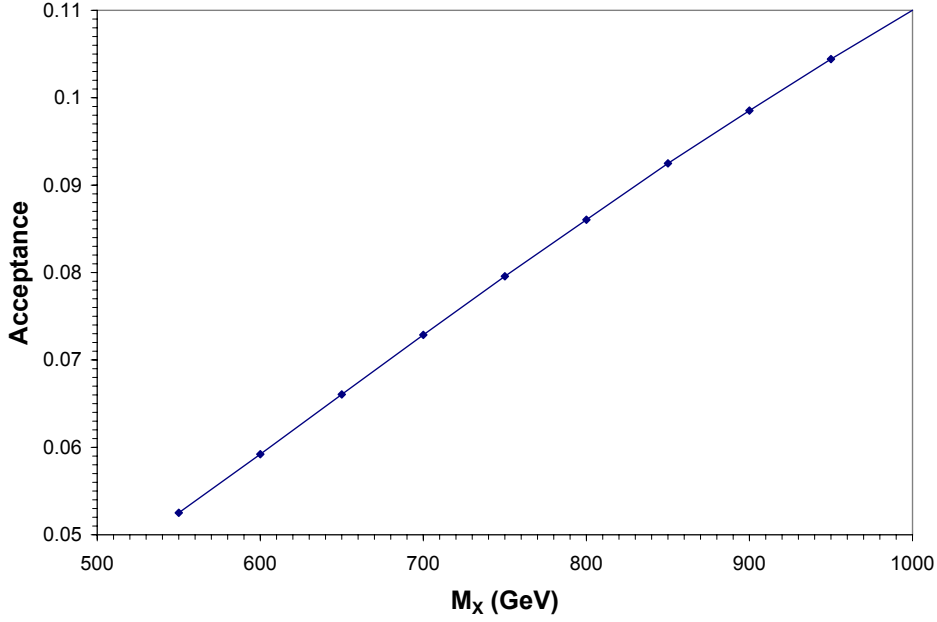}
\caption{Plot of the kinematic and geometric acceptance rate,
which is the fraction of $pp\rightarrow X\to ZZ\to 4l$ events that
satisfy the imposed jet and leptonic cuts.} \label{acceptance}
\end{figure}

 For this gold-plated
signal, one can impose a further set of cuts~\cite{Bagger:1995mk}
on the outgoing leptons:
\begin{itemize}
\item{They reconstruct to two on-shell $Z$'s} \item{$|y(l)| <2.5$}
\item{$p_T (l) >40\gev$} \item{$p_T(Z) > {1\over
4}\sqrt{M^2(ZZ)-4M_Z ^2}$} \item{$M(ZZ) >500\gev$}
\end{itemize}

In our analysis we integrate over the phase space of $pp\to X\to
ZZ\to 4l$ events to determine the total kinematic and geometric
acceptance rate of these cuts. This is defined to be the fraction
of $pp\rightarrow X\to ZZ\to 4l$ events that satisfy the imposed
jet and leptonic kinematic and geometric cuts.  The results are
plotted in fig.~\ref{acceptance}, which shows that about $5\% - 10\%$
of the signal events passes these cuts in the interesting range of
$X$ boson mass.

For the background, after imposing these cuts, one finds that in
the mass range of interest
($M_X \sim 500-1000\gev$),
less than one background event survives in each $50\gev$
bin
with 100 $\xfb^{-1}$ of integrated
luminosity~\cite{Bagger:1995mk}. Detection of this process can
therefore be achieved with 10 signal events in a $50\gev$ bin
centered on $M_{ZZ}$.

\begin{figure}[tb]
\centering
\includegraphics[width=15cm]{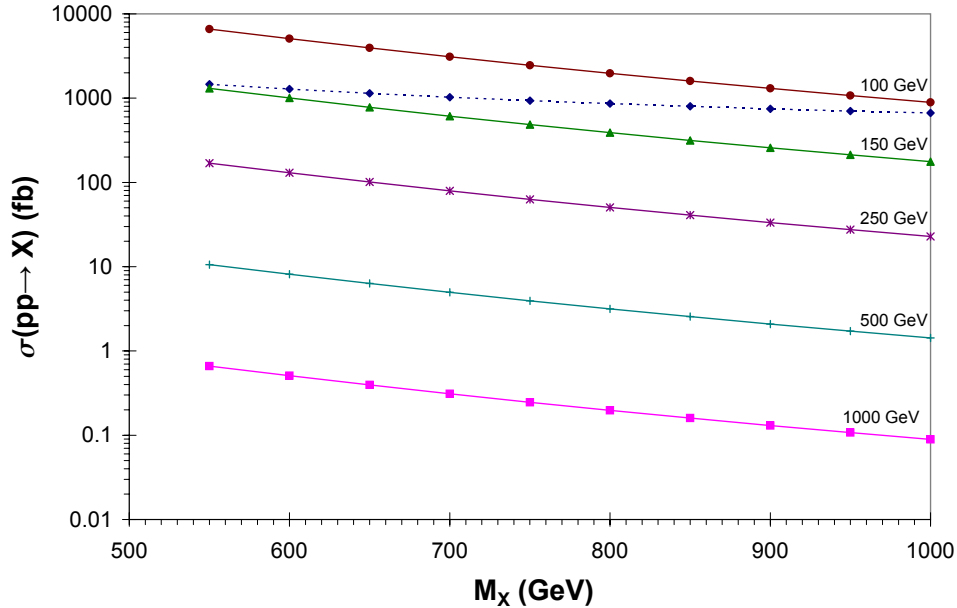}
\caption{Plot of $\sigma(pp\rightarrow X)$ at
$\sqrt{s_{pp}}=14\tev$ LHC as a function of $M_X$ for various
$\Lambda_X$.  The dashed line corresponds to the cross-section
required for detection at LHC in the $X \rightarrow ZZ \rightarrow
4l$ decay channel using the standard leptonic and jet cuts
associated with this gold-plated vector boson fusion channel,
discussed in text.} \label{X signal}
\end{figure}

In fig.~\ref{X signal} we plot the total cross section
$\sigma(pp\to X)$ at
the LHC as a function of the $X$ boson mass, $M_X$. The various
solid lines in the plot correspond to different choices of
$\Lambda_X$.
The dashed line in fig.~\ref{X signal}
shows the required $pp\to X$ production cross-section
in order to find a 10 event signal in a $50\gev$ bin centered on
$M_{ZZ}$ in $100\xfb^{-1}$ in the $4l$ channel.
For $M_X$ in the range $500-1000\gev$, discovery can be made if $\Lambda_X \sim
100-150\gev$.

For any viable model, the fermion which run in the loop must 
be massive (to avoid the appearance of SM chiral exotics).  This 
implies that $U(1)_X$ gauge symmetry must be 
broken \cite{Anastasopoulos:2006da}.  The symmetry breaking effects 
may provide additional signals for the $U(1)_X$ gauge boson.

\section*{Conclusions}

We have considered a physics scenario motivated by intersecting
brane models, in which there are hidden $U(1)$ gauge groups under
which the SM particles are uncharged.
However, there can exist exotic matter, including non-vectorlike
matter, that couples to both SM and hidden sector gauge groups.
The resulting loop diagrams, along with tree-level
higher-dimension couplings arising from the Green-Schwarz anomaly
cancellation mechanism, generate an effective vertex that couples
the hidden-sector gauge boson to two electroweak gauge bosons.

We defined $\Lambda_X$ as the mass-scale that suppresses these
higher-dimensional couplings, and found that in the $X$ boson mass
range of our study ($M_X \sim 500-1000\tev$), LHC could detect
this new physics through $pp\rightarrow X \rightarrow ZZ
\rightarrow 4l$ processes provided $\Lambda_X \sim 100-150\gev$.
Although such a low $\Lambda_X$ is below the naive theory
anticipations, it is important to remember that the contributions
of vectorlike matter can be considerable, and the exponentially
growing Hagedorn spectrum of states could put these values of
$\Lambda_X$ within reach.

One should note that our estimates of detection likelihood are
based only on the gold-plated decay channel $X \rightarrow ZZ
\rightarrow 4l$.  Although this mode combines a clean signal with
a well-studied and highly suppressed background, it does suffer
from small branching fractions.  Other decay modes (for example,
$ZZ\rightarrow ll+2j$, $WW\rightarrow 2l+2\nu$, or $\gamma Z$
decay) could collectively provide even better detection prospects.

For example, an especially interesting and unique signal in this
context is the decay of $X\to\gamma Z \rightarrow \gamma ll$. ${
\Gamma_{\gamma Z} \over \Gamma_{ZZ} } \sim 0.3$ implies a smaller
rate of $\gamma Z$ intermediate states than $ZZ$, but the small
branching fraction of $ZZ \rightarrow 4l$, along with the
cleanness of a $\gamma$ signal make $X\rightarrow \gamma Z$ an
important contributing mode for study.  There has been some work
on $\gamma ll$ signatures in the context of Higgs
searches~\cite{gammall}, but none apparently in conjunction with
vector boson fusion production cuts.  One would expect that
appropriate cuts would also reduce the background for this signal
to negligible levels, though a definitive statement would require
a detailed background analysis which is beyond the scope of this
letter.  A comprehensive search strategy over all $X$ boson decay
chains would be the ideal approach.
This will be left for future work.

We conclude by noting
that the couplings of the
hidden sector $U(1)_X$ bosons to both SM fermions and SM gauge
bosons depend on the details of the hidden sector. Although this
fact makes it difficult to predict precisely how the exotic states
will couple to SM states, the sensitivity gives one a chance to
study and determine the dynamics of the hidden sector should these
exotics appear at the LHC.

\section*{Acknowledgements}
We gratefully acknowledge D. Berenstein, B. Dutta, S. Kachru, E.
Kiritsis and C.-P. Yuan for useful discussions. This work is
supported in part by the Department of Energy, the National
Science Foundation (NSF grants PHY-0314712 and PHY-0555575), and the Michigan
Center for Theoretical Physics. The work of AR is supported in
part by NSF  grant No.~PHY--0354993 and NSF  grant
No.~PHY--0653656.


\begin{thebibliography}{99}

\bibitem{IBM}
  R.~Blumenhagen, L.~Goerlich, B.~Kors and D.~Lust,
  JHEP {\bf 0010}, 006 (2000)
  [hep-th/0007024];
  C.~Angelantonj, I.~Antoniadis, E.~Dudas and A.~Sagnotti,
  Phys.\ Lett.\ B {\bf 489}, 223 (2000)
  [hep-th/0007090];
  R.~Blumenhagen, L.~Goerlich, B.~Kors and D.~Lust,
  Fortsch.\ Phys.\  {\bf 49}, 591 (2001)
  [hep-th/0010198];
  G.~Aldazabal, S.~Franco, L.~E.~Ibanez, R.~Rabadan and A.~M.~Uranga,
  J.\ Math.\ Phys.\  {\bf 42}, 3103 (2001)
  [hep-th/0011073];
  G.~Aldazabal, S.~Franco, L.~E.~Ibanez, R.~Rabadan and A.~M.~Uranga,
  JHEP {\bf 0102}, 047 (2001)
  [hep-ph/0011132];
  M.~Cvetic, G.~Shiu and A.~M.~Uranga,
  Nucl.\ Phys.\ B {\bf 615}, 3 (2001)
  [hep-th/0107166];
  M.~Cvetic, P.~Langacker and G.~Shiu,
  Nucl.\ Phys.\ B {\bf 642}, 139 (2002)
  [hep-th/0206115];

\bibitem{IBM2}
  R.~Blumenhagen, V.~Braun, B.~Kors and D.~Lust,
  hep-th/0210083;
  A.~M.~Uranga,
  Class.\ Quant.\ Grav.\  {\bf 20}, S373 (2003)
  [hep-th/0301032];
  M.~Cvetic, P.~Langacker, T.~j.~Li and T.~Liu,
  Nucl.\ Phys.\  B {\bf 709}, 241 (2005)
  [hep-th/0407178];
  F.~Marchesano and G.~Shiu,
  Phys.\ Rev.\  D {\bf 71}, 011701 (2005)
  [arXiv:hep-th/0408059].
  M.~Cvetic and T.~Liu,
  Phys.\ Lett.\  B {\bf 610}, 122 (2005)
  [hep-th/0409032];
  F.~Marchesano and G.~Shiu,
  JHEP {\bf 0411}, 041 (2004)
  [arXiv:hep-th/0409132].
  C.~Kokorelis,
  hep-th/0410134;
  J.~Kumar and J.~D.~Wells,
  JHEP {\bf 0509}, 067 (2005)
  [arXiv:hep-th/0506252].
  C.~M.~Chen, T.~Li and D.~V.~Nanopoulos,
  Nucl.\ Phys.\ B {\bf 732}, 224 (2006)
  [hep-th/0509059];
  B.~Dutta and Y.~Mimura,
  Phys.\ Lett.\ B {\bf 633}, 761 (2006)
  [hep-ph/0512171];
  M.~R.~Douglas and W.~Taylor,
  JHEP {\bf 0701}, 031 (2007)
  [hep-th/0606109].

\bibitem{kinmix}
  K.~R.~Dienes, C.~F.~Kolda and J.~March-Russell,
  Nucl.\ Phys.\  B {\bf 492}, 104 (1997)
  [hep-ph/9610479];
  J.~Kumar and J.~D.~Wells,
  Phys.\ Rev.\  D {\bf 74}, 115017 (2006)
  [hep-ph/0606183];
  D.~Feldman, Z.~Liu and P.~Nath,
  JHEP {\bf 0611}, 007 (2006)
  [hep-ph/0606294];
  W.~F.~Chang, J.~N.~Ng and J.~M.~S.~Wu,
  Phys.\ Rev.\  D {\bf 74}, 095005 (2006)
  [hep-ph/0608068], and
   hep-ph/0701254.
  D.~Feldman, Z.~Liu and P.~Nath,
  Phys.\ Rev.\  D {\bf 75}, 115001 (2007)
  [hep-ph/0702123];
  W.~F.~Chang, J.~N.~Ng and J.~M.~S.~Wu,
   0706.2345 [hep-ph].

\bibitem{IBMmixanom}
  E.~Kiritsis and P.~Anastasopoulos,
  JHEP {\bf 0205}, 054 (2002)
  [arXiv:hep-ph/0201295].
  D.~M.~Ghilencea, L.~E.~Ibanez, N.~Irges and F.~Quevedo,
  JHEP {\bf 0208}, 016 (2002)
  [arXiv:hep-ph/0205083].
  C.~Coriano, N.~Irges and E.~Kiritsis,
  Nucl.\ Phys.\  B {\bf 746}, 77 (2006)
  [arXiv:hep-ph/0510332].
  B.~Dutta and J.~Kumar,
  Phys.\ Lett.\  B {\bf 643}, 284 (2006)
  [hep-th/0608188].
  B.~Dutta, J.~Kumar and L.~Leblond,
  hep-th/0703278.

\bibitem{Anastasopoulos:2006cz}
  P.~Anastasopoulos, M.~Bianchi, E.~Dudas and E.~Kiritsis,
  JHEP {\bf 0611}, 057 (2006)
  [hep-th/0605225].

\bibitem{anom}
  C.~Coriano, N.~Irges and S.~Morelli,
  arXiv:hep-ph/0701010.
  C.~Coriano, N.~Irges and S.~Morelli,
  arXiv:hep-ph/0703127.




\bibitem{Berenstein:2006pk}
  D.~Berenstein and S.~Pinansky,
  Phys.\ Rev.\  D {\bf 75}, 095009 (2007)
  [hep-th/0610104].

\bibitem{Anastasopoulos:2006da}
  P.~Anastasopoulos, T.~P.~T.~Dijkstra, E.~Kiritsis and A.~N.~Schellekens,
  Nucl.\ Phys.\  B {\bf 759}, 83 (2006)
  [arXiv:hep-th/0605226].




\bibitem{EVBA}
  G.~L.~Kane, W.~W.~Repko and W.~B.~Rolnick,
  Phys.\ Lett.\  B {\bf 148}, 367 (1984);
  M.~S.~Chanowitz and M.~K.~Gaillard,
  Nucl.\ Phys.\  B {\bf 261}, 379 (1985);
  S.~Dawson,
  Nucl.\ Phys.\  B {\bf 249}, 42 (1985);
  Z.~Kunszt and D.~E.~Soper,
  Nucl.\ Phys.\  B {\bf 296}, 253 (1988).


\bibitem{Han:1992hr}
  T.~Han, G.~Valencia and S.~Willenbrock,
  Phys.\ Rev.\ Lett.\  {\bf 69}, 3274 (1992)
  [hep-ph/9206246].

\bibitem{cteq}
CTEQ Collaboration, CTEQ6 parton distribution functions,
http://www.phys.psu.edu/$\sim$cteq/.

\bibitem{tagged jets}
  R.~N.~Cahn {\it et al.},
  Phys.\ Rev.\  D {\bf 35}, 1626 (1987);
  V.~D.~Barger, T.~Han and R.~J.~N.~Phillips,
  Phys.\ Rev.\  D {\bf 37} (1988) 2005;
  R.~Kleiss and W.~J.~Stirling,
  Phys.\ Lett.\  B {\bf 200}, 193 (1988);
  U.~Baur and E.~W.~N.~Glover,
  Phys.\ Lett.\  B {\bf 252}, 683 (1990);
  V.~D.~Barger {\it et al.},
  Phys.\ Rev.\  D {\bf 44}, 1426 (1991).

\bibitem{Bagger:1995mk}
  J.~Bagger {\it et al.},
  Phys.\ Rev.\  D {\bf 52}, 3878 (1995)
  [hep-ph/9504426].

\bibitem{VBFHiggs}
  D.~L.~Rainwater, M.~Spira and D.~Zeppenfeld,
  hep-ph/0203187.
  D.~Green,
  hep-ex/0309031.

\bibitem{gammall}
  J.~F.~Gunion, G.~L.~Kane and J.~Wudka,
  Nucl.\ Phys.\  B {\bf 299}, 231 (1988).







\end{thebibliography}
\end{document}